\documentclass[a4paper,11pt]{article}

\usepackage{jheppub} 
\usepackage{enumerate}
\usepackage[T1]{fontenc} 

\newcommand{\mc}[1]{\mathcal{#1}}

\title{A Detailed Study of Bogomol'nyi Equations in Two-Dimensional Generalized Maxwell-Higgs Model Using \textit{On-Shell} Method}

\author[a,b,1]{A. N. Atmaja,\note{Corresponding author.}}
\author[c]{H. S. Ramadhan,}
\author[d]{and E. da Hora}

\affiliation[a]{Quantum Science Centre, Department of Physics, Faculty of Science, University of Malaya, 50603 Kuala Lumpur, Malaysia.}
\affiliation[b]{Research Center for Physics, Indonesian Institute of Sciences (LIPI), Kompleks PUSPIPTEK Serpong, Tangerang 15310, Indonesia.}
\affiliation[c]{Departemen Fisika, FMIPA, Universitas Indonesia, Depok, 16424, Indonesia.}
\affiliation[d]{Departemento de F\'{i}sica, Universidade Federal do Maranh\~{a}o, 65080-805, S\~{a}o Lu\'{i}s, Maranh\~{a}o, Brazil.}

\emailAdd{ardian\_n\_a@um.edu.my}
\emailAdd{hramad@ui.ac.id}
\emailAdd{edahora.ufma@gmail.com}

\abstract{We use a recent {\it on-shell} Bogomol'nyi method, developed in~\cite{Atmaja:2014fha}, to construct Bogomol'nyi equations of the two-dimensional generalized Maxwell-Higgs model~\cite{Bazeia:2012uc}. The formalism can generate a large class of Bogomol'nyi equations parametrized by a constant $C_0$. The resulting equations are classified into two types, determined by $C_0=0$ and $C_0\neq0$. We identify that the ones obtained by Bazeia {\it et al}~\cite{Bazeia:2012uc} are of the type $C_0=0$. We also reveal, as in the case of ordinary vortex, that this theory does not admit Bogomol'nyi equations in the Bogomol'nyi-Prasad-Sommerfield limit in its spectrum. However, when the vacuum energy is lifted up by adding some constant to the energy density then the existence of such equation is possible. Another possibility whose energy is equal to the vacuum is also discussed in brief. As a future of the \textit{on-shell} method, we find another new Bogomol'nyi equations, for $C_0\neq0$, which are related to a non-trivial function defined as a difference between energy density of potential term of the scalar field and kinetic term of the gauge field.}

\begin{document} 

\maketitle
\flushbottom


\section{Introduction}
Bogomolnyi method is a smart trick to reduce the second-order Euler-Lagrange equations into the first-order, whose solitonic solutions possess minimum energies~\cite{bogol}. For topologically nontrivial field's vacuum manifold the solutions are stable since at the boundary they map each point in coordinate space with different global minimum of the potential.

So far the Bogomolnyi equations were derived by saturating the lower bound of the corresponding static energy (the so-called {\it off-shell} approach). This method may not always give the Bogomolnyi equations easily, especially when the Lagrangian contains noncanonical terms, as in the case of $k$-defects~\cite{kdefects1, kdefects2, kdefects3, kdefects4, kdefects5, kdefects6, kdefects7}. Recently, two of us~\cite{Atmaja:2014fha} proposed an alternative in obtaining the first-order equations by directly evaluating the Euler-Lagrange equations, later dubbed the {\it on-shell} approach. This formalism reproduces the known Bogomolnyi equations for kinks, vortices, and monopoles, as well as Dirac-Born-Infeld (DBI) kins and vortices. This is a novel result though still preliminary, since it might enable us in constructing BPS (Bogomonlyi-Prasad-Sommerfield) states for general defects. Not only it is interesting in its own right, but also these least-energy solitonic solutions might have different properties from their canonical BPS counterparts. In the context of cosmology this might shed a new light on the dynamics of defects.

Not long time ago one of us~\cite{Bazeia:2012uc} studied topological vortices in the generalized Maxwell-Higgs theory, whose dynamics are controlled by two positive functions in the Lagrangian, $G\left(|\phi|\right)$ and $w\left(|\phi|\right)$. It was shown that, for several choices of $G-$ and $w-$functions, there exist BPS solutions with various topology and energies (that can be greater than the canonical BPS tensions). Soon it was followed by the discovery of prescription for obtaining their analytical BPS vortex solutions~\cite{Casana:2014qfa}. The similar study was also done on generalized BPS monopoles~\cite{casana1, casana2, casana3}\footnote{These are truly remarkable results, since the search for analytic BPS vortex solutions has been notoriously difficult and so far has been futile while the finding of BPS monopole solutions by Prasad and Sommerfield was achieved only after several trials and errors~\cite{Prasad:1975kr}. It is the appearance of $G$ and $w$ functions that, in spite of making the EoM appear more complicated, actually helps in obtaining the suitable solutions that satisfy the boundary conditions.}.

Here in this paper we look for something more modest by following a different route. Our aim is twofold. First, we wish to improve the on-shell method so that it includes noncanonical Lagrangian. Second, by applying it to the generalized Maxwell-Higgs theory we try to construct generators that generates the corresponding Bogomol'nyi equations. It is expected that for arbitrarily positive functions $G\left(|\phi|\right)$ and $w\left(|\phi|\right)$ a large class of first-order Bogomolnyi equations (and their solutions) can be obtained.

\section{Improved Version of On-shell Method}
\label{sec: improved}
The effective, one-dimensional, Euler-Lagrange equations (6) in the on-shell method of~\cite{Atmaja:2014fha} are difficult to get since the right hand side of the equations is only allowed to depend on the parameter $r$ and the fields $\phi^a$. It was very fortunate that examples given in~\cite{Atmaja:2014fha} for the non-standard theory, which were the DBI defects, have not suffered from this difficulty. However, it should not happen in general for any theory with non-standard kinetic terms, such as the Generalized Maxwell-Higgs theory discussed in this article. Here, we need to improve the on-shell method such that the right hand side of the effective Euler-Lagrange equations are allowed to depend on first derivative of the fields $\phi^a$. As a simple case, let us consider a theory with the effective degree of freedom is given by $\phi$, in which the effective one dimensional Lagrangian $\mc{L}=\mc{L}(r,\phi,\phi')$ and the Euler-Lagrange equation are given by
\begin{eqnarray}
\label{gen 1-field}
0&=&{\partial\mc{L}\over\partial\phi}-{d\over dr}\left(\partial\mc{L}\over\partial\phi'\right)\nonumber\\ 
0&=&\mc{A}(r,\phi,\phi')-\mc{B}_r(r,\phi,\phi')-\mc{B}_\phi(r,\phi,\phi') \phi'-\mc{B}_{\phi'}(r,\phi,\phi') \phi''\nonumber\\
\phi''&=&{1\over \mc{B}_{\phi'}(r,\phi,\phi')}\left(\mc{A}(r,\phi,\phi')-\mc{B}_r(r,\phi,\phi')-\mc{B}_\phi(r,\phi,\phi') \phi'\right),
\end{eqnarray}
where
\begin{equation}
 \mc{A}={\partial\mc{L}\over\partial\phi},\ \ \ \ \mc{B}_x={\partial\over\partial x}\left(\partial\mc{L}\over\partial\phi'\right),\ \ \ \ x\equiv\left(r,\phi,\phi'\right).
\end{equation}
The Euler-Lagrange equation can be arranged into
\begin{equation}
\label{Alternative1}
 \phi''+f(r,\phi,\phi')\phi'=0.
\end{equation}
We then need to determine what would be the expected function of $f(r,\phi,\phi')$ provided that the left hand side of (\ref{Alternative1}) can be rewritten as
\begin{equation}
 \phi''+f(r,\phi,\phi')\phi'={1\over h}(h\phi')'+\ldots,
\end{equation}
where $h\equiv h(r,\phi)$. Now, since $h'={\partial h\over \partial r}+{\partial h\over \partial\phi}\phi'$, it yields that the function $f$ must be of the form
\begin{equation}
 f(r,\phi,\phi')={1\over h}{\partial h\over \partial r}+{1\over h}{\partial h\over \partial\phi} \phi'+(\mbox{non-linear terms in }\phi').
\end{equation}
We keep the linear terms of $f$, in $\phi'$, in the left hand side of (\ref{Alternative1}) and move the non-linear terms to the right hand side of (\ref{Alternative1}). The Bogomol'nyi equation is then given by $h(r,\phi)\phi'=X(\phi)$, while the constraint equation is now
\begin{equation}
 {X'\over h}=g(r,\phi,\phi'), 
\end{equation}
where $g$ contains all remaining non-linear terms coming from $f$. Notice that upon substituting the Bogomol'nyi equation into (\ref{Alternative1}), we can get back the form of effective Euler-Lagrange equation as in the equation (6) of~\cite{Atmaja:2014fha}.

For multiple fields theory\footnote{Here, we follow the conventions in~\cite{Atmaja:2014fha} for $N_\phi-$fields theory.}, generalization of the above procedures are more involved. As such, for each field $\phi^a$, the effective one dimensional Euler-Lagrange equations are
\begin{eqnarray}
\label{gen N-field}
 0&=&\mc{A}^a(r,\phi,\phi')-\mc{B}^a_r(r,\phi,\phi')-\sum_{b}\mc{B}^a_{\phi^b}(r,\phi,\phi') {\phi^b}'-\sum_{b}\mc{B}^a_{{\phi^b}'}(r,\phi,\phi') {\phi^b}'',\nonumber\\
{\phi^a}''&=&{1\over \mc{B}^a_{{\phi^a}'}(r,\phi,\phi')}\left(\mc{A}^a(r,\phi,\phi')-\mc{B}^a_r(r,\phi,\phi')-\sum_{b\neq a}\mc{B}^a_{\phi^b}(r,\phi,\phi') {\phi^b}'-\sum_{b\neq a}\mc{B}^a_{{\phi^b}'}(r,\phi,\phi') {\phi^b}''\right),\nonumber\\
\end{eqnarray}
where
\begin{equation}
 \mc{A}^a={\partial\mc{L}\over\partial\phi^a},\ \ \ \ \mc{B}^a_x={\partial\over\partial x}\left(\partial\mc{L}\over\partial{\phi^a}'\right),\ \ \ \ x\equiv\left(r,\phi^b,{\phi^b}'\right),\ \ \ \ b=1,\ldots,N_\phi.
\end{equation}
One should notice that the Euler-Lagrange equations are linear in $\phi''$. Taking the same procedures as in the case of a single field theory, we may write the Euler-Lagrange equation, for each $\phi^a$, as
\begin{equation}
 {\phi^a}''+f^a(r,\phi,\phi'){\phi^a}'=g^a(r,\phi,\phi')+\sum_{b\neq a}k_{ab}(r,\phi,\phi')\left[{\phi^b}''+f^b(r,\phi,\phi'){\phi^b}'\right],
\end{equation}
where $f$ is linear function in ${\phi}'$. To have the Bogomol'nyi equations, the function $f^b$ must be of the form 
\begin{equation}
f^b(r,\phi,\phi')={1\over h^b}{\partial h^b\over \partial r}+{1\over h^b}\sum_{c}{\partial h^b\over \partial\phi^c}{\phi^c}',\ \ \ \qquad c=1,\ldots,N_\phi,
\end{equation}
where $h^b=h^b(r,\phi)$. The Bogomol'nyi equations then are given by 
\begin{equation}
 h^b(r,\phi){\phi^b}'=X^b(\phi)
\end{equation}
and the constraint equations are
\begin{equation}
 {{X^a}'\over h^a}=g^a(r,\phi,\phi')+\sum_{b\neq a}k_{ab}(r,\phi,\phi'){{X^b}'\over h^b}.
\end{equation}

As in~\cite{Atmaja:2014fha}, the topological charge can directly be obtained by inserting the Bogomol'nyi equations into the energy functional. We shall obtain, in general,
\begin{equation}
\label{eq: topocharge}
dQ=\sum_{a}F[X^{a}(\phi)]{\phi^a}',
\end{equation}
where $F[X^a(\phi)]$ is a general functional of $X^a(\phi)$ whose form depends on the actual kinetic form of the Lagrangian. In particular, for canonical case $F[X^a(\phi)]=X^a(\phi)$. Its integral becomes
\begin{eqnarray}
\label{eq: Energibps}
E_{BPS}&=&\int dQ,\nonumber\\
&=& Q(r=\infty)-Q(r=0).
\end{eqnarray} 

\section{Generalized Maxwell-Higgs Model}
As an example of application of the prescription above, let us now consider a generalized Maxwell-Higgs theory described by the following $(1+2)$-dimensional Lagrangian density~\cite{Bazeia:2012uc}
\begin{equation}
\label{Lagrangian}
 \mc{L}_G=-{1\over 4}G(|\phi|)F_{\mu\nu}F^{\mu\nu}+w(|\phi|) |D_\mu\phi|^2-V(|\phi|),
\end{equation}
where $F_{\mu\nu}=\partial_\mu A_\nu-\partial_\nu A_\mu$, $D_\mu\phi=\partial_\mu+i e A_\mu\phi$, and the Minkowskian metric is $\eta^{\mu\nu}\equiv diag(+,-,-)$. Here, we take the gauge coupling $e$ and the vacuum expectation value $v$ of the scalar field to be real and positive. The functions $G\left(|\phi|\right)$ and $w\left(|\phi|\right)$ are constrained to be positive and depend explicitly only on the Higgs field amplitude, $|\phi|$, but not on its derivative\footnote{The case for field-derivative-dependent functions will be addressed in the forthcoming publication.}. In this article, we will consider a static solitonic object, in particular topological vortices, in which all the fields are static. Furthermore, we will consider the spatial part of the action and write it in terms of the spherical coordinates.

We chose a temporal gauge $A^0=0$ and the static fields ansatz
\begin{equation}
 \phi= v\ g(r) e^{i n\theta},\qquad\qquad \textbf{A}=-{a(r)-n \over e\ r}\hat{\theta},
\end{equation}
where $(r,\theta)$ is the polar coordinates and $n=\pm1,\pm2,\ldots$ is an integer winding number. Notice that the Lagrangian is invariant under two-dimensional rotation and an abelian gauge transformation, $SO(2)\times U(1)$. The ansatz for the Higgs field is chosen to be invariant under subgroup of this symmetry which is the $SO(2)$ rotational transformation with a particular choice of $U(1)$ gauge transformation, that cancels the two-dimensional rotation. It is guaranteed that the solutions of the effective equations of motion, derived by using this ansatz, are also the solutions of the full equation of motions~\cite{Manton:2004tk}. 

Using these ansatz, the static energy, proportional to the static action, can be simply written as
\begin{equation}
\label{energydensity}
E=2\pi\int dr\ r\left({G\over2e^2}\left({a'\over r}\right)^2+v^2 w \left(g'^2+{g^2a^2\over r^2}\right)+V\right).
\end{equation}
The Euler-Lagrange equations, or equations of motion, derived from the above static energy are
\begin{equation}
 G{d^2a \over dr^2}+\left({dG\over dr}-{G\over r}\right){da\over dr}= 2 e^2 v^2 g^2 a w,
\end{equation}
and
\begin{equation}
 w\left({d^2g\over dr^2}+{1\over r}{dg\over dr}-{a^2g \over r^2}\right)-{1\over 4 v^2}\left({1\over e r}{da\over dr}\right)^2{dG\over dg}= {1\over 2 v^2}{dV\over dg}-{1\over 2}\left(\left(dg\over dr\right)^2-{g^2a^2\over r^2}\right){dw\over dg}.
\end{equation}
The vacuum solution of the above theory (\ref{Lagrangian}) is related to the solution in which $A_\mu=0$ and $\phi=v$. For the case of topological vortex, we consider the case in which $v\neq0$. For topological vortex solutions, we require the fields $a$ and $g$ to behave asymptotically, near the origin and the boundary, as follows
\begin{align}
\label{boundary conditions}
 a(r\to0)=n, \ \ \ g(r\to0)=0, \nonumber \\
 a(r\to\infty)=0, \ \ \ g(r\to\infty)=1.
\end{align}
How fast the functions $a$ and $g$ approaching their asymptotic values, namely the next leading order terms, is determined by the Bogomol'nyi equations and the explicit form of $G$, $w$, and $V$; as such, the static energy (\ref{energydensity}) is finite.

\section{Bogomol'nyi Equations}

In order to obtain the Bogomol'nyi equations, following the prescription in section~\ref{sec: improved}, we rewrite the Euler-Lagrange equations into
\begin{equation}
\label{eq: first EL}
 {r\over G} {d\over dr}\left({G\over r}{da\over dr}\right)={2\over G} e^2 v^2 g^2 a w,
\end{equation}
and
\begin{equation}
\label{eq: second EL}
 {1\over rw^{1/2}}{d\over dr}\left(r w^{1/2}{dg\over dr}\right)={1\over 4wv^2 e^2 G^2}\left({G\over r}{da\over dr}\right)^2+{a^2 g\over r^2}+{1\over 2 v^2}{dV\over dg}+{g^2a^2\over 2 r^2 w}{dw\over dg}.
\end{equation}
The first term on the right hand side of equation (\ref{eq: second EL}) contains first derivative of field $a$, $a'(r)$. It can be turned into a non-derivative fields dependence by using the Bogomol'nyi equations as we will show later in detail. Now, let us introduce some auxiliary fields into the Euler-Lagrange equations as follows
\begin{equation}
  r {d\over dr}\left({G\over r}{da\over dr}-X\right)+r{dX\over dr}=2 e^2 v^2 g^2 a w,
\end{equation}
and
\begin{equation}
{w^{1/2}\over r}{d\over dr}\left(r w^{1/2}{dg\over dr}-Y\right)+{w^{1/2}\over r}{dY\over dr}={1\over 4v^2 e^2 G^2}\left({G\over r}{da\over dr}\right)^2{dG\over dg}+{a^2w g\over r^2}+{1\over 2 v^2}{dV\over dg}+{g^2a^2\over 2 r^2}{dw\over dg},
\end{equation}
where $X$ and $Y$ are the auxiliary functions that depend only on the fields $a$ and $g$, but not their derivatives, and do not depend explicitly on $r$.
From these equations, we can extract the Bogomol'nyi equations which are
\begin{equation}
\label{eq: BPS-1}
 {G\over r}{da\over dr}-X=0
\end{equation}
and
\begin{equation}
\label{eq: BPS-2}
 r w^{1/2}{dg\over dr}-Y=0.
\end{equation}
The Bogomol'nyi equations are supplemented by the constraint equations
 \begin{equation}
 \label{eq: Constraint-1}
  r{dX\over dr}=2 e^2 v^2 g^2 a w,
 \end{equation}
and
 \begin{equation}
  \label{eq: Constraint-2}
  {w^{1/2}\over r}{dY\over dr}={X^2\over 4v^2 e^2 G^2}{dG\over dg}+{a^2w g\over r^2}+{1\over 2 v^2}{dV\over dg}+{g^2a^2\over 2 r^2}{dw\over dg}.
 \end{equation}
Notice that we have substituted the first term on the right hand side of the constraint equation (\ref{eq: Constraint-2}) by using the Bogomol'nyi equation (\ref{eq: BPS-1}). Substituting further the Bogomol'nyi equations into the constraint equations yields
\begin{equation}
 {\partial X\over\partial g} {Y\over r w^{1/2}}+{\partial X\over\partial a} {r X\over G}={2\over r} e^2 v^2 g^2 a w,
\end{equation}
 and
 \begin{equation}
  {\partial Y\over\partial g} {Y\over r w^{1/2}}+{\partial Y\over\partial a} {r X\over G}={r\over w^{1/2}}\left({X^2\over 4v^2 e^2 G^2}{dG\over dg}+{a^2w g\over r^2}+{1\over 2 v^2}{dV\over dg}+{g^2a^2\over 2 r^2}{dw\over dg}\right).
 \end{equation}

Next, we solve those constraint equations by dividing each of them into terms that depend on the explicit power of $r$. Solving those terms independently, this process yields several equations:
\begin{equation}
\label{eq: reduced constraint-1}
 {\partial X\over\partial a}=0,\ \ \ \ \ {\partial X\over\partial g} {Y\over w^{1/2}}=2 e^2 v^2 g^2 a w,
\end{equation}
\begin{equation}
\label{eq: reduced constraint-2}
 {\partial Y\over\partial g} Y =a^2 w g+{g^2a^2\over 2}{dw\over dg},\ \ \ \ \ {\partial Y\over\partial a} {X\over G}={X^2\over 4v^2 e^2 G^2 w^{1/2}}{dG\over dg}+{1\over 2 v^2 w^{1/2}}{dV\over dg}.
\end{equation}
The problem is now reduced to finding the auxiliary functions, $X$ and $Y$, which solve the above (constraint) equations. The fist equation in (\ref{eq: reduced constraint-1}) implies that $X$ is independent of $a$. The general solution for $Y$ can be obtained by solving the first equation in (\ref{eq: reduced constraint-2}) which is given by $Y^2(g,a)=a^2 g^2 w+C_0(a)$, where $C_0$ is an arbitrary function of $a$. However, for nontrivial solutions, the second equation in (\ref{eq: reduced constraint-1}) restricts the function $C_0\propto a^2$. In general, we may write the solution for $Y$ to be $Y^2(g,a)=a^2\left(g^2 w+C_0\right)$, where now $C_0$ is just a constant. Since the first equation in (\ref{eq: reduced constraint-1}) gives $X\equiv X(g)$, all the auxiliary functions are essentially separable functions. Writing all the auxiliary functions to be separable,
\begin{equation}
 X(g,a)=X_g(g) X_a(a),\ \ \ \ \ Y(g,a)=Y_g(g) Y_a(a),
\end{equation}
without loss of generality we can take $X_a=1$, $Y_a=a$, and $Y_g^2=g^2w+C_0$. Using $Y_g=\pm \sqrt{g^2 w+C_0}$, we obtain from the second equation in (\ref{eq: reduced constraint-1})
\begin{equation}
X_g=\pm e^2v^2\left(2\int dg {g^2 w^{3/2} \over \sqrt{g^2w+C_0}}+C_1\right),
\end{equation}
where $C_1$ is an integration constant. Therefore we obtain that the Bogomol'nyi equation (\ref{eq: BPS-1}) depends on functions $w$ and $G$, while the Bogomol'nyi equation (\ref{eq: BPS-2}) depends only on function $w$.

It will be useful later to define functions
\begin{equation}
\label{defined RS}
 R(g)={X_g\over G},\ \ \ \ \ S(g)={Y_g\over w^{1/2}}.
\end{equation}
Using the previously obtained functions: $X_g$ and $Y_g$, we are left with only one constraint equation, the second equation in (\ref{eq: BPS-2}), which in terms of functions $S$ and $R$ is simply written as
\begin{equation}
\label{eq: generator V}
 V'=2v^2  w R S-{R^2 \over 2 e^2}G'.
\end{equation}
From now on, we will use $'\equiv {\partial\over \partial g}$ if it not defined explicitly. The Bogomol'nyi equations can simply be rewritten as follows
\begin{equation}
\label{BPS RS}
 r {dg \over dr}= a~S,\ \ \ \ \ {1\over r}{da \over dr}= R .
\end{equation}
So, we can say that the equations in (\ref{defined RS}) generate the Bogomol'nyi equations in (\ref{BPS RS}) for the generalized Maxwell-Higgs model (\ref{Lagrangian}) once we fix the functions: $w$ and $G$, and the constants: $C_0$ and $C_1$, while the constraint equation (\ref{eq: generator V}) determines the form of potential $V$ once we know all these functions and constants. At first sight, the constraint equation (\ref{eq: generator V}) is different from the standard one obtained in~\cite{Bazeia:2012uc} which, in our conventions, can be written as
\begin{equation}
\label{eq: Ampere}
 \left(\sqrt{G~V\over 2}\right)'= e v^2 w g.
\end{equation}
However, we will show later in the next section that the constraint equation (\ref{eq: Ampere}) of~\cite{Bazeia:2012uc} is a particular case of our constraint equation (\ref{eq: generator V}).


\subsection{Bogomol'nyi equations for \texorpdfstring{$C_0=0$}{C0 = 0}}
In this subsection, we consider a particular simple class of solutions. This class of solutions is provided by taking $C_0=0$, for which we obtain $S=\pm q$ and
 \begin{equation}
 \label{eq: sol Bazeia}
  X_g=\pm e^2 v^2 \left(\int d(g^2)~w+C_1\right).
 \end{equation}
It is tempted to expect from the above integral that $w\equiv w(g^2)$ which happens to be the case in all Bogomol'nyi equations of~\cite{Bazeia:2012uc}. One can also check that all functions of $w,G,$ and $V$ in each Bogomol'nyi equations of~\cite{Bazeia:2012uc} are solutions to the constraint equation (\ref{eq: generator V}). In this case, the Bogomol'nyi equations can be simply written as
\begin{eqnarray}
 r{dg \over dr}&=&\pm a~g,\\
 {1 \over r}{da \over dr}&=&\pm{e^2 v^2\over G}\left[\int d(g^2)~w+C_1\right],
\end{eqnarray}
and the constraint equation (\ref{eq: generator V}) now becomes
\begin{equation}
 V'=\pm 2v^2  w R g-{R^2 \over 2 e^2}G'.
\end{equation}
Using the fact that $X_g'=(RG)'=\pm 2 e^2 v^2 g w$, the constraint equation can be rewritten as
\begin{equation}
\label{eq: generator V C_0=0}
 V'={1\over e^2} R (GR)'-{R^2 \over 2 e^2}G'={R^2\over 2e^2}G'+{RG \over e^2} R'.
\end{equation}
The solution to this differential equation is
\begin{equation}
\label{potential C_0=0}
 V={1\over 2e^2}R^2G+\mbox{constant}={e^2v^4 \over 2 G}\left[\int d(g^2) w+C_1\right]^2+\mbox{constant}.
\end{equation}
Here, this constant can actually be set to zero by shifting the potential $V$ in the action. Furthermore, we will see later that by imposing a condition that the energy of the vortex to be finite, this constant is forced to be zero. In this case, it turns out that the potential (\ref{potential C_0=0}) also solves the constraint equation (\ref{eq: Ampere}), and thus it is the same as the constraint equation in~\cite{Bazeia:2012uc}. The potentials obtained in~\cite{Bazeia:2012uc} can be derived simply by using the constraint (\ref{potential C_0=0}) with a particular choice of the functions and parameters:
\begin{enumerate}[(a).]
\label{list of BPS equations}
 \item Standard Maxwell-Higgs model\\
 $G=1;\ \ \ w=1;\ \ \ C_1=-1\ \ \ \longrightarrow\ \ \ V={e^2 v^4\over 2}(1-g^2)^2$.
 \item $G={(g^2+3)^2\over g^2};\ \ \ w=2(g^2+1);\ \ \ C_1=-3\ \ \ \longrightarrow\ \ \ V=g^2{e^2 v^4\over 2}(1-g^2)^2$.
 \item $G=(g^2+1)^2;\ \ \ w=2g^2;\ \ \ C_1=-1\ \ \ \longrightarrow\ \ \ V={e^2 v^4\over 2}(1-g^2)^2$.
 \item $G={k^2 \over 2e^2v^2g^2};\ \ \ w=1;\ \ \ C_1=-1\ \ \ \longrightarrow\ \ \ V={e^4 v^6\over k^2}g^2(1-g^2)^2$.
\end{enumerate}


\subsubsection*{Flat potential}
A slight advantage of our constraint equation (\ref{eq: generator V C_0=0}) is that the potential $V$ can be safely taken to be zero. Unlike the one in~\cite{Bazeia:2012uc}, or equation (\ref{eq: Ampere}), setting $V=0$ will not give us a solution. In the limit of the coupling at which the potential $V=0$,  the Bogomol'nyi equations would corresponds to the BPS equations for vortex; this is similar to the case of BPS (Bogomol'nyi-Prasad-Sommerfield) monopole~\cite{Prasad:1975kr}. This limit is known as the BPS limit which is essential if one wants to construct the supersymmetric version of the theory. In this case the solution for $G$ is given by
\begin{equation}
\label{flat solution}
 G=C_2^2 e^4 v^4 \left(\int d(g^2)~w+C_1\right)^2 \longrightarrow C_2^2R^2G=1,
\end{equation}
where $C_2$ is an non-zero integration constant related to the non-zero constant in (\ref{potential C_0=0}). Although the constraint (\ref{eq: Ampere}) is not suitable for the case of $V=0$, the solution (\ref{flat solution}) can actually be obtained from it by setting the potential to be constant $V={1 \over 2 e^2 C^2_2}$. This is related to the fact, as we will discuss in the next section, that the finiteness energy requires a shift in the potential by a constant. Nevertheless, the Bogomol'nyi, or to be precise BPS, equations now become
\begin{eqnarray}
\label{flatBPSeq1}
 {dg \over dr}&=&\pm {ag \over r},\nonumber\\
 {da \over dr}&=&\pm{r\over C_2 \sqrt{G}}.
\end{eqnarray}
Here, the function $G$ depends on the function $w$ and the constants ($C_1$ and $C_2$). We present some of the examples, with $C_2=1$, as follows
\begin{itemize}
 \item $w=1;\ \ \ C_1=-1\ \ \ \longrightarrow\ \ \ G= e^4 v^4(g^2-1)^2$.
 \item $w=2g^2;\ \ \ C_1=-1\ \ \ \longrightarrow\ \ \ G= e^4 v^4(g^4-1)^2$.
 \item $w=2(g^2+1);\ \ \ C_1=-3\ \ \ \longrightarrow\ \ \ G= e^4 v^4(g^2-1)^2(g^2+3)^2$.
\end{itemize}
 
Later, we will find that all of the above examples turn out to give infinite energy. This can be seen due to the presence of singularity of the corresponding BPS equations near the boundary. As an example, consider the configuration (c) above in which $w=2(g^2+1)$ and $C_1=-3$ gives $G= e^4 v^4(g^2-1)^2(g^2+3)^2$. The BPS equations are
\begin{align}
g'&=\pm{ag\over r},\notag\\
a'&=\pm{r\over e^{2}v^{2}\left(g^{2}-1\right)\left(g^2+3\right)}.
\end{align}
The second equation blows up at infinity, since $g(r\rightarrow\infty)\rightarrow 1$. 
On the other hand, there should be many possibilities of $G(g)$ such that it satisfies the boundary conditions. For example, we can take 
\begin{equation}
G={e^4v^4\over (1-g^2)^{2}}.
\end{equation}
This can be obtained by taking\footnote{This choice opens up a possibility that $G$ and $w$ can take up rational-form functions.}
\begin{equation}
w={1\over \left(1-g^2\right)^2},\ \ \ C_1=0.
\end{equation}
It is amusing that the combination of $G$ and $w$ above, when inserted into the equations (\ref{flatBPSeq1}), produces  precisely the equations for ordinary BPS Maxwell-Higgs vortices (up to some overall constants),
\begin{eqnarray}
\label{flatBPSsol}
g'&=&\pm{ag\over r},\nonumber\\
{a\over r}&=&\pm{\left(1-g^2\right)\over e^2\nu^2}.
\end{eqnarray} 
Since we know that BPS vortices exist, so do these {\it flat potential} BPS generalized Maxwell-Higgs vortices. However, there is a subtle here that the functions $w$ and $G$ now can be singular near the boundary, or $g\to1$.

\subsection{Bogomol'nyi equations for \texorpdfstring{$C_0\neq 0$}{C0 != 0}}

For a general case, we can rewrite the constraint equation (\ref{eq: generator V}) in terms of $R$ and $S$ as follows
\begin{equation}
\label{eq: sol V}
 V'={R^2\over e^2}\left({S^2\over g^2}-{1\over 2}\right)G'+{RG \over e^2} {S^2\over g^2} R'.
\end{equation}
Unlike the $C_0=0$ case, the right hand side of the contraint equation above is more complicated and it is very difficult to write it as a total derivative of some functions and hence difficult to find the solution. However, we may try to follow what we did as in the $C_0=0$ case and write the constraint (\ref{eq: sol V}) simply as
\begin{equation}
 2e^2V'=\left(R^2G^2\right) \left(1\over G\right)'+{S^2\over g^2 G}\left(R^2G^2\right)'.
\end{equation}
To have a total derivative, we are tempted to identify
\begin{equation}
 {1\over G}+C_3={S^2\over g^2}{1\over G},
\end{equation}
where $C_3$ is just a constant which we can just add to the constraint equation above by shifting $\left(1\over G\right)'\to\left({1\over G}+C_3\right)'$. The value of $C_3$ needs to be non-zero otherwise it would not be consistent with $C_0\neq0$ since $S^2=g^2$. With this identification, we obtain that
\begin{equation}
 G={1\over g^2 w}{C_0\over C_3}.
\end{equation}
This is consistent with the $C_0=0$ solutions, in which we have to take $C_3=0$ in order for $G$ to be non-trivial. However, this is a little bit peculiar because $G$ dependence of $w$ is in contradiction with the $C_0=0$ solutions. We might expect that $G$ is still independent of $w$, or arbitrary, for this more general case in which the constant $C_0$ is arbitrary. It turns out that this solution can not lead to the finite energy solution as discussed in the next section.

Although the constraint equation (\ref{eq: sol V}) does not seem to have a solution, let us write explicitly the Bogomol'nyi equations:
\begin{align}
 {dg \over dr}&=\pm {a\over r}\sqrt{g^2w+C_0 \over w},\\
 {da \over dr}&=\pm e^2 v^2 {r\over G}\left(2\int dg {g^2 w^{3/2} \over \sqrt{g^2w+C_0}}+C_1\right).
\end{align}
Even if we are able to find solutions for the constraint equation (\ref{eq: sol V}), it is not guaranteed that those solutions will have finite energy. We will see later that there are some possibilities in which the solutions to the constraint equation (\ref{eq: sol V}) would give a finite energy.



\section{Static Energy}
\label{sec: energy}
The static energy, given by formula~(\ref{energydensity}), can be rewritten into a nicer form by using Bogomol'nyi equations (\ref{BPS RS}),
\begin{equation}
 E_{Sol}=2\pi\int \left(\left({G R\over 2e^2 }+{V\over R}\right)da+{v^2w a\over S}\left(S^2+g^2\right)dg\right),
\end{equation}
which is defined as the static energy of vortex. We could define a function $Q\equiv2\pi a\left({G R\over 2e^2 }+{V\over R}\right)$, such that $E_{Sol}=\int dQ$, if we could solve
\begin{equation}
 V'=v^2 w(S^2+g^2){R\over S}-{R^2 \over 2 e^2}G'+\left({V \over R}-{RG \over 2 e^2}\right)R'.
\end{equation}
Substituting this equation into the constraint (\ref{eq: sol V}) yields
\begin{equation}
\label{New constraint 1}
 \left({RG \over 2}-{e^2 V \over R}\right)R'=e^2v^2{w R\over S}(g^2-S^2),
\end{equation}
or we can also write
\begin{equation}
\label{New constraint 2}
 \left({e^2 V \over G}-{R^2 \over 2}\right)G'=2e^2v^2{g^2 w \over R S}\left({e^2 V \over G}-{R^2 \over 2}{S^2 \over g^2}\right).
\end{equation}

Now let us see if the vortices have finite energy using the Derrick's Theorem~\cite{Derrick:1964,Manton:2004tk}. We can write the scaled static energy of (\ref{energydensity}) to be
\begin{eqnarray}
 E(\lambda)&=&\lambda^2 E_{gauge}+E_{scalar}+{1\over \lambda^2}E_{pot},\nonumber \\
 E_{gauge}&=&\int d^2x~{G\over 2e^2 r^2}\left(da\over dr\right)^2,\ \ \ \ \ \ \ E_{pot}=\int d^2x~ V,\nonumber \\
 E_{scalar}&=&\int d^2x\left(v^2 w\left(dg\over dr\right)^2+v^2a^2g^2 {w\over r^2}\right),
\end{eqnarray}
 where $0<\lambda<\infty$ is the scale factor. There is a stationary point if we vary the $E(\lambda)$ over $\lambda$ at which is finite and positive. It means there are some vortices with finite energy. Furthermore, the virial theorem requires $E_{gauge}=E_{pot}$. Consider a simple case which both energy densities are equal pointwise. After substituting the Bogomol'nyi equations (\ref{BPS RS}), it yields that 
\begin{equation}
\label{finite energy C_0=0}
 V ={R^2G \over 2 e^2},
\end{equation}
Substituting this into the equation (\ref{New constraint 1}), or (\ref{New constraint 2}), implies that $S^2=g^2$, or it means $C_0=0$. Therefore if we assume that the energy  can be written as an integral over a form $dQ$ then the $C_0\neq 0$ vortices will have infinite energy. Even if we do not use this assumption and just use the equation (\ref{finite energy C_0=0}), we can show that
\begin{equation}
 V'=-{R^2 \over 2 e^2}G'+2v^2 g^2 w {R\over S}
\end{equation}
by taking a first derivative of the equation (\ref{finite energy C_0=0}) over $g$ and using the Bogomol'nyi equations (\ref{BPS RS}).
This is equal to the constraint (\ref{eq: sol V}) providing that $S^2=g^2$, which also concludes that $C_0=0$. Therefore we may ignore the $C_0\neq0$ Bogomol'nyi equations as they are not physical since their energy is infinite. 

\subsection{Finite energy for \texorpdfstring{$C_0=0$}{C0 = 0}}
Notice that the requirement for the energy of the solution to be finite, for the point wise case, forces us to set the constant in (\ref{potential C_0=0}) to be zero. In this case, the static energy (\ref{energydensity}) can be simplified to
\begin{eqnarray}
 E_{Sol}&=&2\pi\int dr~r\left({GR\over e^2}{1\over r}{da\over dr}+2v^2w{g a \over r}{dg\over dr}\right),\nonumber\\
	&=&2\pi\int\left({GR \over e^2}da+ 2v^2w a g~dg\right).
\end{eqnarray}
Recalling that $(RG)'=2e^2v^2 g w$, we can obtain the aforementioned function $Q=2\pi{GR\over e^2}a$. Now, the static energy is simply written as
\begin{equation}
 E_{Sol}=Q(r\to\infty)-Q(r\to 0).
\end{equation}
Using formula (\ref{eq: sol Bazeia}), it yields\footnote{Notice that for any polynomial $w$-function, $w(g)\sim g^m$ with $m\geq 0$, the $\lim_{r\to 0}\int w~d(g^2)$ always yields zero. In this case the topological charge is solely determined by the constant $C_{1}$. On the other hand, we can also easily construct a rational w-function, say 
\begin{equation}
\label{fracW}
w={1\over(g^2+1)^2}. 
\end{equation}
This function is positive and regular at the origin, whose (indefinite) integral gives 
\begin{equation}
\int w~d(g^2)=-{1\over(g^2+1)}.
\end{equation}
The limit then yields -1. In this particular case, the charge would depend on $(C_1-1)$.}
\begin{equation}
 E_{Sol}=\left|2\pi v^2 n\left(\lim_{r\to 0}\int w~d(g^2)+C_1\right)\right|.
\end{equation}
Here, we have assumed that $Q(r\to\infty)=0$ and hence $GR(r\to\infty)=O(1)$. In other words, we assume that $GR$ is not singular near the boundary. This can be shown to be satisfied in general by writing $GR=\sqrt{2e^2 VG}$, using the equation (\ref{finite energy C_0=0}). Recalling that near the boundary, the potential $V$ approaches the vacuum solution, in which $V=0$, then it only requires that $G(r\to\infty)=O(1)$. The finiteness of energy also requires $\lim_{r\to 0}\int w~d(g^2)$, or $\sqrt{VG}(r\to 0)$, to be finite. Since all these functions ($w, G$ and $V$) are functions of $g$, we may rewrite it as $\left.\int w~d(g^2)\right|_{g=0}$, or $\sqrt{VG}(g=0)$, to be finite.

The static energy can be proportional to the topological charge $Q_{Top}=2\pi v^2 |n|$ as such $E_{Sol}=C_Q Q_{Top}$, where $C_Q\geq0$. In the case of $C_Q=1$, we obtain that
\begin{equation}
\lim_{r\to 0}\int w~d(g^2)+C_1=\pm 1,
\end{equation}
which means the static energy equal to the standard vortex. For example for $w=1$, we have $C_1=\pm 1$. In the list of examples in \ref{list of BPS equations}, the (a), (c), and (d) are of this type. If $C_Q> 1$ then the static energy is higher than the standard vortex, $E_{Sol}>Q_{Top}$, and they are determined by \begin{equation}
 \left|\lim_{r\to 0}\int w~d(g^2)+C_1\right|>1.
\end{equation}
The example (b) in \ref{list of BPS equations} is in this type in which the static energy $E_{Sol}=3 Q_{Top}$. There are also some interesting Bogomol'nyi equations in which $C_Q<1$, or $E_{BPS}<Q_{Top}$, and the condition is given by
\begin{equation}
 \left|\lim_{r\to 0}\int w~d(g^2)+C_1\right|<1.
\end{equation}
The most interesting of one is when $C_Q=0$, or $E_{BPS}=0$, with a condition
\begin{equation}
 C_1=-\lim_{r\to 0}\int w~d(g^2).
\end{equation}
This raises a question, do Bogomol'nyi solutions with zero energy exist? A rigorous answer needs a rigorous proof. In this paper we do not attempt to answer it. We just note that if we choose the following set of functions and parameter
\begin{equation}
w=2g^2-1,\ \ \ G=1,\ \ \ C_1=0,
\end{equation} 
we can end up with the following Bogomolnyi equation
\begin{equation}
{a'\over r}=\pm e^2\nu^2g^2\left(1-g^2\right),
\end{equation}
whose potential is $V={e^2\nu^4\over 2}g^4\left(1-g^2\right)^2$, an $S^0$ surrounded by an $S^1$ vacuum topology. The equation satisfies both regularity at the origin and  finiteness of energy. Due to the vacuum manifold, this is an example of nontopological soliton discussed in~\cite{Bazeia:2012uc, epl}\footnote{For Bogomol'nyi topological solitons we need potential whose vacuum manifold is nontrivial. For example if $w=1$ then it yields $C_1=0$. Now we can set $G=g^4/(1-g^2)^2$ such that the theory still has the standard symmetry breaking Higgs potential $V={1\over 2} e^2v^4 (1-g^2)^2$. However these functions do not satisfy the near origin condition for the Bogomol'nyi equation; {\it i.e.,}
\begin{equation}
{a'\over r}=\pm e^2\nu^2{\left(1-g^2\right)^2\over g^2}
\end{equation}
is singular at the origin.}.

\subsubsection*{Flat potential}
As we mentioned previously, finiteness in the static energy requires the constant in the constraint equation (\ref{potential C_0=0}) to be zero, or $2e^2 V=R^2 G$. Taking $V=0$ is not possible in this case and might cause the resulting energy to be infinite. Nevertheless, let us just ignore the requirement for finite energy and allow the potential $V=0$. The static energy in this case can be written as
\begin{eqnarray}
 E_{Sol}&=&2\pi\int \left[{RG\over 2e^2} da+ 2v^2w a g~dg\right],\nonumber\\
	&=&\int dQ-{\pi\over e^2}\int RG~da.
\end{eqnarray}
Substituting the BPS equations (\ref{flatBPSeq1}), we obtain
\begin{eqnarray}
\label{energyflat}
 E_{Sol}&=&{2\pi \over e^2\left|C_2\right|} \left.{a\sqrt{G}}\right|^{r\to\infty}_{r=0}-\lim_{r\to\infty}{\pi \over 2 e^2 C^2_2} r^2.
\end{eqnarray}
Indeed, we find that the static energy is infinite which comes from the last term on the right hand of equation (\ref{energyflat}). This infinity can be removed by adding a constant potential in the action\footnote{Since we do not, at the moment, couple the theory with gravity, adding a constant potential does not change the physics.}. The constant potential needed to remove this infinity is equal to the potential computed using equation (\ref{finite energy C_0=0}) with a given solution for $G$ is (\ref{flat solution}). Therefore if we take the potential to be non-zero constant in the first place, we will have no problem in taking the finite energy equation (\ref{finite energy C_0=0}), and thus the static energy will be finite. 

The first term on the right hand side of equation (\ref{energyflat})depends on $a\sqrt{G}$ at the boundaries. As we mentioned previously, there is a subtlety in function $G$ if we impose regularity on the Bogomol'nyi equations (\ref{flatBPSeq1}). To have Bogomol'nyi equations (\ref{flatBPSeq1}) that respect appropriate boundary conditions, $G\sim\left(1-g^2\right)^{-2m}$, for some positive integer $m$. Although $G$ is infinite near the boundary, by taking appropriate leading order of function $a$ as such it is going to zero faster than $1/\sqrt{G}$, we could obtain the static energy which is
\begin{equation}
E_{Sol}=\left|2\pi v^2n\lim_{r\rightarrow 0}a\sqrt{G}\right|.
\end{equation}
For our case in equation (\ref{flatBPSsol}), it yields
\begin{equation}
E_{Sol}=\left|{2\pi v^2n\over C_2}\right|.
\end{equation}
It is interesting that the arbitrary choice of $C_2$ results in different value of $E_{Sol}$.

\subsection{Finite energy for \texorpdfstring{$C_0\neq 0$}{C0 != 0}}

From the previous discussion, it is clear that the finite-energy equation (\ref{finite energy C_0=0}) strongly restricts the constant $C_0=0$. Therefore the Bogomol'nyi equations for $C_0\neq0$ would not give a finite static energy of the vortex. However, we should recall that the equation (\ref{finite energy C_0=0}) is not the only result of the Derrick's theorem followed by the virial theorem. There is another, more general, result of the virial theorem that the finiteness of energy requires
\begin{equation}
\label{finite energy C_0 not 0}
\int^\infty_{0} dr ~r \left({R^2G \over 2e^2}-V\right)=0,
\end{equation}
in which the integrand is non-zero pointwise. Up to now, we do not know how to substitute the definite integral equation (\ref{finite energy C_0 not 0}) into the constraint equation (\ref{eq: generator V}). What we can do is that we can try to rewrite the constraint equation (\ref{eq: generator V}) to be the following
\begin{equation}
 \left(V-{R^2G \over 2e^2}\right)'={C_0 R \over e^2 g^2 w} \left(R G\right)'.
\end{equation}
Using $(RG)'=2e^2v^2g^2w/S$, it can be simplified further to
\begin{equation}
\label{modified gen constraint}
 \left(V-{R^2G \over 2e^2}\right)'=2C_0v^2 {R\over S}.
\end{equation}
One can see that if $C_0=0$ then the left hand side of equation (\ref{modified gen constraint}) must be some constant. However, if this constant is non-zero then the integral equation (\ref{finite energy C_0 not 0}) can not be satisfied. Therefore the constant must be zero and indeed it is consistent with the finite energy equation (\ref{finite energy C_0=0}). Now, if $C_0\neq0$ then the left hand side of equation (\ref{modified gen constraint}) must be some function. Suppose we define $f(g)=V-{R^2G \over 2e^2}$ is a function solely depends on $g$. To have a finite energy, using equation (\ref{finite energy C_0 not 0}), this function must satisfy
\begin{align}
 \int^\infty_{0} dr ~r f(g(r))=0.
\end{align}
Indeed, there are many solutions for $f$, in terms of parameter $r$, that satisfy this condition. One of them is given by the special Laguerre functions with the following integral~\cite{Schaum}
\begin{align}
 \int^\infty_{0} dr~r~e^{-r}L_n(r)&=0,\ \ \ \ \ \ L_n(r)=e^r {d^n\over dr^n}\left(r^n e^{-r}\right),
\end{align}
where $L_n$ is the Laguerre functions for $n>1$. Substituting the function $f$, in terms of $r$, into the constrain equation (\ref{modified gen constraint}), and exploiting the Bogomol'nyi equations (\ref{BPS RS}), it yields solution for $a$ as follows  
\begin{align}
 a^2={1\over 2C_0v^2} \int dr~r^2f'(r)+C_a,
\end{align}
where now $'\equiv{d\over dr}$ and $C_a$ is an integration constant. Finding suitable function for $f(r)$, that satisfy the boundary conditions (\ref{boundary conditions}), might give us the explicit form of functions $w(g)$ and later also $G(g)$.

\section{Summary}
The main purpose of this article is to show how the \textit{on-shell} method, developed in~\cite{Atmaja:2014fha}, can be used to find the Bogomol'nyi equations of the generalized Maxwell-Higgs theory in three-dimensional spacetime~\cite{Bazeia:2012uc}. In particular, we improved the \textit{on-shell} method to allow the terms in the equations of motion, that would later be identified as the constraint equations, to depend on the derivative of the fields. The improvement is necessary to tackle a particular type of theory such as the one considered in this article. This might open some possibilities to improve and modify the \textit{on-shell} method in obtaining the Bogomol'nyi equations of the other non-standard theories.

In the case of the generalized Maxwell-Higgs theory, we found that the Bogomol'nyi equations can be classified into two types which are parametrized by a constant $C_0$. The first type is for $C_0=0$ in which we obtained the standard Bogomol'nyi equations as shown in~\cite{Bazeia:2012uc,Casana:2014qfa}. An advantage of using the \textit{on-shell} method is that we obtained the constraint equation (\ref{eq: generator V C_0=0}) that can be applied for the case of zero potential. Although it turns out that the resulting energy is infinite, we were able to show that the static energy could be finite by adding an appropriate non-zero constant to the potential. We also discussed possibilities for the existence of vortices with the energy is equal to the vacuum. From what we know, this has not been discussed in the literature so far and it might be interesting to study the physical properties of this vortices compared to the vacuum.

The second new type Bogomol'nyi equations, that we found here, is when $C_0\neq0$. These equations are relatively new and we do think they could not be obtained easily using the standard \textit{off-shell}, or Hamiltonian, method. It turned out that these equations are related to the difference between the energy density of potential term of the scalar field and kinetic term of the gauge field which is given by a non-trivial function $f$. If the function $f$ is a constant then the requirement for finite energy vortex forces this constant to be zero and hence gives us back the the first type of Bogomol'nyi equations, $C_0=0$. The requirement for finite energy vortex on the Bogomol'nyi equations of the second type, $C_0=0$, restrict further the function $f$ such that its integral over whole two-dimensional space is zero. Here, we do not attempt to find the explicit expressions of the Bogomol'nyi equations of the second type since they will be discussed in the future work.


\acknowledgments
A.N.A acknowledges University of Malaya for the support through the University of Malaya Research Grant (UMRG) Programme RP006C-13AFR and RP012D-13AFR. H.S.R acknowledges support from University of Indonesia through Research Cluster Grant on ``Non-perturbative phenomena in nuclear astrophysics and cosmology" No.~1709/H2.R12/HKP. 05.00/2014.

\end{document}